\documentclass{article}
\usepackage{amssymb}

\usepackage{graphicx}

\begin{document}

\title{Multiple steady states and the form of response functions to antigen 
in a model for the initiation of T cell activation}

\author{Alan D. Rendall\\
Institut f\"ur Mathematik\\
Johannes Gutenberg-Universit\"at\\
Staudingerweg 9\\
D-55099 Mainz\\
Germany\\
\\
and\\
\\
Eduardo D. Sontag\\
Department of Mathematics and the Center for Quantitative Biology\\
Rutgers University\\
Piscataway, NJ 08854\\
USA}

\date{}

\maketitle

\begin{abstract}
The aim of this paper is to study the qualitative behaviour 
predicted by a mathematical model for the initial stage of T cell activation. 
The state variables in the model are the concentrations of phosphorylation 
states of the T cell receptor complex and the phosphatase SHP-1 in the 
cell. It is shown that these quantities cannot approach zero and that the 
model possesses more than one positive steady state for certain values of the 
parameters. It can also exhibit damped oscillations. It is proved that the 
chemical concentration which represents the degree of activation of the cell,
that of the maximally phosphorylated form of the T cell receptor complex, is 
in general a non-monotone function of the activating signal. In particular 
there are cases where there is a value of the dissociation constant of the 
ligand from the receptor which produces an optimal activation of the T cell.  
In this way the results of certain simulations in the literature have 
been confirmed rigorously and some important features which had not previously 
been seen have been discovered.
\end{abstract}

\section{Introduction}

In humans and other vertebrates the immune system is of crucial importance for
protecting an individual from dangers such as pathogens, toxins and cancer.
(For background information on immunology we refer to \cite{murphy12}.)
The central players in the immune system are the white blood cells (leukocytes)
and it is important that these cells be able to distinguish between dangerous
substances and host tissues. This is often referred to as the distinction 
between non-self and self. A failure to combat dangerous substances may lead 
to infectious diseases becoming life-threatening. On the other hand, if the 
immune system attacks host tissues this can lead to autoimmune disease. The 
task of discrimination is complicated. An important element of the process of 
distinction between self and non-self is the activity of the class of leukocytes
called T cells. An individual T cell is supposed to recognize a particular
substance (antigen) and take suitable action if that substance is dangerous.
Recognition is based on the binding of the antigen to a molecule on the T cell
surface, the T cell receptor (TCR). It is believed that the most important 
aspect of this process is the time the antigen remains bound before being 
released (the dissociation time), an idea which has been called the 'lifetime 
dogma' \cite{feinerman08a}. When it recognizes its antigen the T cell changes 
its behaviour and is said to be activated. In what follows we study a 
mathematical model for what happens in the first few minutes after a T cell 
recognizes its antigen. 

In \cite{altanbonnet05} Altan-Bonnet and Germain introduced a model for the 
initial stage of T cell activation. Simulations using this model gave results
which fitted a number of experimental findings. On the other hand it was too 
elaborate to be readily accessible to a mathematical analysis of its dynamics. 
In \cite{francois13} the authors introduced a radically simplified version of 
the model of \cite{altanbonnet05}. The new model includes the essential 
explanatory power of the old one while being much more transparent and 
tractable for analytical investigation. It also made some new predictions 
which were confirmed experimentally. In \cite{francois13} a number of
interesting analytical calculations were performed but the 
mathematical conclusions which can be drawn from these were not worked out
in detail.

The aim of the present paper is to obtain results about the qualitative 
behaviour of solutions of the model of \cite{francois13} which are as 
general as possible. In Section \ref{def} the model is defined and some
of its basic properties are derived. The model describes a situation where
both an agonist (the antigen which should be recognized) and an 
antagonist (a competing antigen) are present. Section \ref{steady} is concerned 
with the number of steady states and their stability. After some general 
results have been derived, the discussion turns to more detailed properties
of the solutions in the case that the antagonist is absent and treats cases 
where the number $N$ of phosphorylation sites included in the model is small. 
In particular it is shown that when $N=3$ there are parameters for which 
three positive steady states exist (Theorem 1). A numerical calculation 
reveals that for a specific choice of these parameters two of the steady 
states are stable while the third is a saddle. For $N\le 2$ there is a unique
steady state and in the case $N=1$ it is proved to be globally asymptotically
stable. There are parameter values for which the approach to this steady state
is oscillatory.

The qualitative behaviour of the steady state concentration of the maximally 
phosphorylated state, which expresses the degree of activation of the T cell, 
as a function of the antigen concentration and the dissociation time, is 
investigated in the case where only the agonist is present in Section 
\ref{response}. Let us consider the function $f(L_1,\nu_1)$, which expresses
the degree of activation in terms of the parameters $L_1$ (concentration
of agonist ligand) and $\nu_1$ (reaction rate for the dissociation of the 
ligand from the receptor, i.e. the reciprocal of the dissociation time).
It is shown that the dependence exhibits certain types of non-monotone
behaviour in some cases. The results obtained include both rigorous results
on general features of the function $f$ (Theorem 2) and simulations which 
reveal more detailed features. In particular it is found that are values of the 
parameters in the model for which the function $f$ has a maximum as a function
of $\nu_1$ for fixed $L_1$. In other words, there is a value of the 
dissociation time which is optimal for T cell activation. Thus the model
studied here is able to reproduce this fact which has been experimentally 
observed \cite{lever16}.

The analysis of the response function is extended to cover the effects of the 
antagonist in Section \ref{antagonist}. The last section is devoted to 
conclusions and an outlook.

\section{Definition of the model}\label{def} 

In the introduction it was stated that a T cell recognizes an antigen. In 
more detail the molecule concerned is a peptide (a small protein) which is 
bound to a host molecule called an MHC molecule. Thus we talk about a pMHC 
complex as the object to be recognized. In the model of \cite{francois13} two 
types of pMHC complexes are considered. The first, called an agonist, 
represents the case where the antigen comes from a pathogen and should 
activate the T cell. The second, called an antagonist, represents the case of 
a self-antigen, which should not activate the T cell. Detection takes place 
through the binding of a pMHC complex to the T cell receptor.
When this happens certain proteins associated to the T cell receptor are 
phosphorylated, i.e. phosphate groups become attached to them. For simplicity we
will describe this by saying that the receptor-pMHC complex is phosphorylated.

\begin{figure}[ht]
\begin{center}
\includegraphics[scale=0.5]{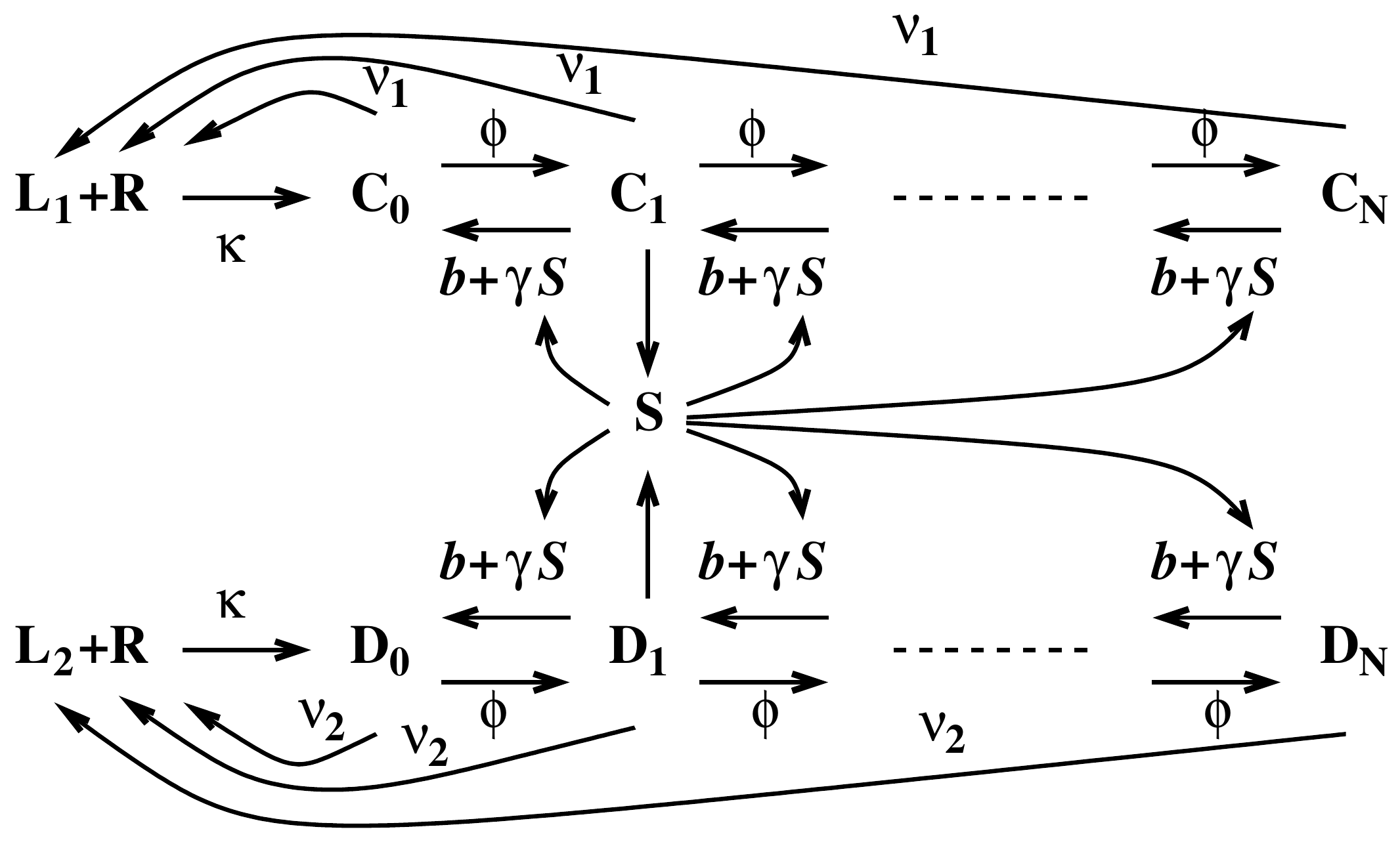}
\end{center}
\caption{The model considered in this paper.  The species $R$ represents the T 
cell
receptor, and $L_1$ and $L_2$ are the two ligands, i.e. the agonist and
antagonist.  The species $C_0$ represents unphosphorylated complexes of the T
cell receptor with the agonist, and the $C_j$'s are the $j$-phosphorylated
complexes.  The $D_j$'s are the analogous complexes for the antagonist.
The phosphatase SHP-1 provides a negative feedback, and is represented by $S$.
The different reactions represent receptor complex phosphorylation with 
rate constant $\phi$ and dephosphorylation with rate constant $b$, as well
as receptor complex dephosphorylation by $S$ with rate constant $\gamma$ 
and dissociation rate constants $\nu_1$ and $\nu_2$. 
Antigens bind to $R$ with rate constant $\kappa$, and $S$ is activated by
the singly phosphorylated complexes with rate constant $\alpha$ and
deactivated with rate constant $\beta$.
}
\label{fig:reactionnetwork}
\end{figure}

The reaction network for the model of \cite{francois13} is shown in
Figure~\ref{fig:reactionnetwork}.
The state variables will now be listed. The 
concentration of 
unphosphorylated complexes of the T cell receptor with the agonist will be 
denoted by $C_0$ and the concentration of unphosphorylated complexes of the T 
cell receptor with the antagonist will be denoted by $D_0$. $C_j$ and $D_j$ 
are the corresponding quantities for the case of $j$ phosphorylations, up to a 
maximum value $N$. The specific value of $N$ has little influence in what 
follows but it may be worth to note that a biologically reasonable value of 
$N$ could be as large as 20 while in the model of \cite{altanbonnet05} we
have $N=9$. $R$, $L_1$ and $L_2$ are the total concentrations of receptors and 
the two ligands, i.e. the agonist and antagonist. Another important element of 
the system is SHP-1. This substance is a phosphatase which means that when
active it can remove phosphate groups from the receptor-pMHC complex. It 
contributes a negative feedback loop to the system. $S$ is the concentration 
of active SHP-1. The receptor complexes are subject to phosphorylation with 
rate constant $\phi$ and dephosphorylation with rate constant $b$. They are 
also dephosphorylated by SHP-1 with rate constant $\gamma $ and dissociate 
with rate constants $\nu_1$ and $\nu_2$. Antigens bind to the receptor with 
rate constant $\kappa$. SHP-1 is activated by the singly phosphorylated 
complexes with rate constant $\alpha$ and deactivated with rate constant 
$\beta$. All the rate constants are assumed positive. $S_T$ is the total 
concentration of SHP-1. It is assumed that all reactions exhibit mass action
kinetics and this leads to the following system of equations
\begin{eqnarray}
&&\dot S=\alpha (C_1+D_1)(S_T-S)-\beta S,\label{phen1}\\
&&\dot C_0=\kappa (L_1-\sum_{j=0}^NC_j)(R-\sum_{j=0}^N (C_j+D_j))
+(b+\gamma S)C_1-(\phi+\nu_1)C_0,\label{phen2}\\
&&\dot C_j=\phi C_{j-1}+(b+\gamma S)C_{j+1}-(\phi+b+\gamma S+\nu_1)C_j,
\ \ 1\le j\le N-1,\label{phen3}\\
&&\dot C_N=\phi C_{N-1}-(b+\gamma S+\nu_1)C_N,\label{phen4}\\
&&\dot D_0=\kappa (L_2-\sum_{j=0}^ND_j)(R-\sum_{j=0}^N (C_j+D_j))
+(b+\gamma S)D_1-(\phi+\nu_2)D_0,\label{phen5}\\
&&\dot D_j=\phi D_{j-1}+(b+\gamma S)D_{j+1}-(\phi+b+\gamma S+\nu_2)D_j,
\ \ 1\le j\le N-1,\label{phen6}\\
&&\dot D_N=\phi D_{N-1}-(b+\gamma S+\nu_2)D_N.\label{phen7}
\end{eqnarray}
In a direct formulation of the system as arising from the reaction network 
it is necessary to include the concentrations of free ligands, free
receptors and inactive phosphatase. This extended system has four conservation
laws corresponding to the total amounts of ligands, receptors and phosphatase.
Using these to eliminate the additional variables leads to the system 
(\ref{phen1})-(\ref{phen7}).

The right hand sides of the equations are Lipschitz and so there is a unique 
solution corresponding to each choice of initial data. To have a biologically 
relevant solution the quantities in the extended system should be non-negative.
It is a well-known fact for reaction networks of this type that data for which 
all concentrations are positive give rise to solutions with the same property 
and that data for which all concentrations are non-negative give rise to 
non-negative solutions. In terms of (\ref{phen1})-(\ref{phen7}) this implies 
statements about the positivity of the quantities $S$, $C_j$ and $D_j$ and of 
the differences $S_T-S$, $R-\sum_{j=0}^N(C_j+D_j)$, $L_1-\sum_{j=0}^NC_j$ and 
$L_2-\sum_{j=0}^ND_j$. Let us call the region where all these quantities are 
strictly positive the biologically feasible region. Note that due to the 
conservation laws this region is bounded. Let $\Sigma_1=\sum_{j=0}^NC_j$ and
$\Sigma_2=\sum_{j=0}^NC_j$. Then it follows from (\ref{phen1})-(\ref{phen7})
that
\begin{eqnarray}
&&\dot\Sigma_1=\kappa(L_1-\Sigma_1)(R-\Sigma_1-\Sigma_2)-\nu_1\Sigma_1,
\label{total1}\\
&&\dot\Sigma_2=\kappa(L_2-\Sigma_2)(R-\Sigma_1-\Sigma_2)-\nu_2\Sigma_2.
\label{total2}
\end{eqnarray}

\noindent
{\bf Lemma 1} Consider a solution 
$(S(t),C_0(t),\ldots,C_N(t), D_0(t),\ldots,D_N(t))$ in the closure of 
the biologically feasible region. Then if 
$(S^*,C_0^*,\ldots,C_N^*,D_0^*,\ldots,D_N^*)$ is an $\omega$-limit point 
of this solution it
is also in the biologically feasible region. In particular, any
steady state is in the biologically feasible region.

\noindent
{\bf Proof} If $S^*=S_T$ we can consider the solution starting at that point 
at some time $t_0$. Since the $\omega$ limit set of a given solution is 
invariant the solution under consideration lies entirely in the $\omega$-limit
set of the original solution. In particular, it is contained in the closure of
the biologically feasible region. The solution starting at the point with
$S^*=S_T$ satisfies $\dot S(t_0)<0$ and therefore the inequality $S(t)>S_T$ for 
$t$ slightly less than $t_0$, a contradiction. In a similar way equation
(\ref{total1}) implies that $\sum_{j=0}^NC^*_j$ cannot attain the value $L_1$
and equation (\ref{total2}) implies that $\sum_{j=0}^ND^*_j$ cannot attain the 
value $L_2$. Summing (\ref{total1}) and (\ref{total2}) shows that 
$\sum_{j=0}^NC^*_j+\sum_{j=0}^ND^*_j$ cannot attain the value $R$.

Note next that $C_0$ cannot be zero at an $\omega$-limit point. For if it
were zero at such a point we could consider the solution passing through that 
point at a time $t_0$. The equation (\ref{phen2}) would imply that 
$\dot C_0(t_0)>0$ and that $C_0(t)<0$ for $t$ slightly less than $t_0$, a
contradiction. Once the positivity of $C_0$ has been proved we can use 
equation (\ref{phen3}) with $j=1$ to show that $C_1$ cannot be zero at an 
$\omega$-limit point. This in turn allows us to prove using (\ref{phen1}) that 
$S$ can never be zero at an $\omega$-limit point.
In a similar way it can be concluded successively that $C_2,\ldots,C_N$ and 
$D_0,\ldots,D_N$ are positive at any $\omega$-limit point of a non-negative 
solution. This concludes the proof of the lemma.

The fact that all $\omega$-limit points of solutions in the closure of the
biologically feasible region are in the biologically feasible region together
with the fact that the closure of that region is compact implies that the 
infimum of the distance of a given solution to the boundary in the limit 
$t\to\infty$ is strictly positive. When this last property holds the system
is said to be persistent \cite{butler86}. Note in addition that the 
closure of the biologically feasible region is convex and hence homeomorphic 
to a closed ball in a Euclidean space. It follows from the Brouwer fixed 
point theorem that a steady state exists (cf. \cite{hale09}, Chapter I, 
Theorem 8.2). Since steady states on 
the boundary have already been excluded we can conclude that there is at least 
one steady state in the biologically feasible region for any fixed choice of 
parameters. That this is the case was assumed implicitly in \cite{francois13}.

\section{Multiplicity of steady states}\label{steady}

A question not addressed in \cite{francois13} is whether there might exist 
more than one positive steady state for a fixed choice of parameters. In this 
section it will be shown that for some values of $N$ and the reaction 
constants this is the case. The aim is to find any parameter values with this 
property while not worrying for the moment how biologically relevant this 
choice of parameters is. Let $f_1$ and $f_2$ denote the right hand sides of
equations (\ref{total1}) and (\ref{total2}). Then 
$\frac{\partial f_1}{\partial\Sigma_2}$ and 
$\frac{\partial f_2}{\partial\Sigma_1}$ are negative and hence the system
(\ref{total1})-(\ref{total2}) is competitive. It follows that every solution
of this system converges to a steady state as $t\to\infty$ \cite{hirsch05}. 

A steady state $(\Sigma_1^*,\Sigma_2^*)$ of (\ref{total1})-(\ref{total2}) 
satisfies the equations
\begin{eqnarray}
&&\kappa(L_1-\Sigma_1^*)(R-\Sigma_1^*-\Sigma_2^*)-\nu_1\Sigma_1^*=0
\label{totalstat1},\\
&&\kappa(L_2-\Sigma_2^*)(R-\Sigma_1^*-\Sigma_2^*)-\nu_2\Sigma_2^*=0
\label{totalstat2}.
\end{eqnarray}
We can solve for $\Sigma_1^*$ and $\Sigma_2^*$ as functions of 
$\Sigma_1^*+\Sigma_2^*$.
\begin{eqnarray}
&&\Sigma_1^*=\frac{\kappa L_1(R-\Sigma_1^*-\Sigma_2^*)}
{\kappa (R-\Sigma_1^*-\Sigma_2^*)+\nu_1},\\
&&\Sigma_2^*=\frac{\kappa L_2(R-\Sigma_1^*-\Sigma_2^*)}
{\kappa (R-\Sigma_1^*-\Sigma_2^*)+\nu_2}.
\end{eqnarray}
Hence
\begin{equation}
\kappa (L_1+L_2-\Sigma_1^*-\Sigma_2^*)
=\frac{\kappa L_1\nu_1}{\kappa (R-\Sigma_1^*-\Sigma_2^*)+\nu_1}
+\frac{\kappa L_2\nu_2}{\kappa (R-\Sigma_1^*-\Sigma_2^*)+\nu_2}.
\end{equation}
The function of $\Sigma_1^*+\Sigma_2^*$ on the left hand side of this equation 
is decreasing on the interval $[0,L_1+L_2]$. The function on the right hand 
side is increasing on the interval $[0,R]$. Their graphs can intersect in at 
most one point. We already know that they must intersect since a positive
steady state of the full system exists. That they intersect can also be seen 
directly. For in all cases the left hand side is greater than the right hand
side for $\Sigma_1^*+\Sigma_2^*=0$ and the opposite inequality holds for
$\Sigma_1^*+\Sigma_2^*=\min\{L_1+L_2,R\}$. Thus the equation has a unique 
solution for $\Sigma_1^*+\Sigma_2^*$ in the interval $[0,\min\{L_1+L_2,R\}]$
From this it is possible to compute values of $\Sigma_1^*$ and 
$\Sigma_2^*$ which solve (\ref{totalstat1}) and  (\ref{totalstat2})
and lie in the intervals $[0,\min\{L_1,R\}]$ and $[0,\min\{L_2,R\}]$, 
respectively. The quantities $\Sigma_1^*$ and $\Sigma_2^*$ are functions of the 
parameters $R$, $L_1$, $L_2$, $\kappa$, $\nu_1$ and $\nu_2$.

It can be concluded that the solution passing through an $\omega$-limit point 
of a solution of the original system satisfies a simplified system containing 
$\Sigma_1^*$ and $\Sigma_2^*$ as parameters. $C_0$ and $D_0$ can be eliminated 
from this system in favour of the other $C_j$ and $D_j$. The result is
\begin{eqnarray}
&&\dot S=\alpha (C_1+D_1)(S_T-S)-\beta S,\label{phenlim1}\\
&&\dot C_1=\phi\Sigma_1^*+(b+\gamma S-\phi)C_2
-(2\phi+b+\gamma S+\nu_1)C_1-\phi\sum_{j=3}^N C_j\label{phenlim2}\\
&&\dot C_j=\phi C_{j-1}+(b+\gamma S)C_{j+1}-(\phi+b+\gamma S+\nu_1)C_j,
\ \ 2\le j\le N-1,\label{phenlim3}\\
&&\dot C_N=\phi C_{N-1}-(b+\gamma S+\nu_1)C_N,\label{phenlim4}\\
&&\dot D_1=\phi\Sigma_2^*+(b+\gamma S-\phi)D_2
-(2\phi+b+\gamma S+\nu_2)D_1-\phi\sum_{j=3}^N D_j,\label{phenlim5}\\
&&\dot D_j=\phi D_{j-1}+(b+\gamma S)D_{j+1}-(\phi+b+\gamma S+\nu_2)D_j,
\ \ 2\le j\le N-1,\label{phenlim6}\\
&&\dot D_N=\phi D_{N-1}-(b+\gamma S+\nu_2)D_N.\label{phenlim7}
\end{eqnarray}
This form of the equations is valid for $N\ge 3$. In the case $N=2$ it is still
correct if it is taken into account that the condition $2\le j\le N-1$ is never
satisfied so that the equations containing that condition are absent. The sum 
from $j=3$ to $N$ is zero in that case. The case $N=1$ is exceptional from the 
point of the notation.  

In order to get more information we will restrict in the remainder of this
section to what we call the agonist-only case. This is obtained from the system 
(\ref{phen1})-(\ref{phen7}) by setting $L_2$ and the $D_i$ to zero. There
is a corresponding limiting system, which is obtained from
(\ref{phenlim1})-(\ref{phenlim7}) by setting $\Sigma_2^*$ and the $D_i$ to
zero. In this case we write $\Sigma^*$ instead of $\Sigma_1^*$ for brevity.
Consider the limiting system in the agonist-only case with $N=1$. This is 
\begin{eqnarray}
&&\dot S=\alpha C_1 (S_T-S)-\beta S\label{red1s},\\
&&\dot C_1=\phi\Sigma^*-(\phi+b+\gamma S+\nu_1)C_1\label{red1c}.
\end{eqnarray}
Solving the equation $\dot S=0$ for $C_1$ and substituting the result
into the equation $\dot C_1=0$ gives the quadratic equation 
\begin{equation}\label{S1}
\beta\gamma S^2+[\beta (\phi+b+\nu_1)+\alpha\phi\Sigma^*]S
-\alpha\phi\Sigma^* S_T=0.
\end{equation}
Since the quadratic polynomial has positive leading term and is negative for 
$S=0$ it is clear that it has a unique positive root. It follows from
(\ref{S1}) that this root is less than $S_T$. Equation (\ref{red1c}) implies
that $C_1<\Sigma^*$ at a steady state and so these quantities can be completed 
to a steady state of the original system by defining $C_0=\Sigma^*-C_1$. The 
steady state is unique in this case.

In the case $N=2$ the equations are
\begin{eqnarray}
&&\dot S=\alpha C_1(S_T-S)-\beta S,\\
&&\dot C_1=\phi\Sigma^*-(2\phi+b+\gamma S+\nu_1)C_1+(-\phi+b+\gamma S)C_2,\\
&&\dot C_2=\phi C_1-(b+\gamma S+\nu_1)C_2.
\end{eqnarray}
Proceeding in a manner analogous to what we did in the case $N=1$ it is possible
to get a cubic equation for $S$ in the case $N=2$, which we can write 
schematically in the form $p(S)=\sum_{k=0}^Na_kS^k$. We have
\begin{eqnarray}
&&a_0=-\alpha S_T(b+\nu_1)\phi\Sigma^*,\nonumber\\
&&a_1=\beta [b(\phi+b+\nu_1)+\nu_1(2\phi+b+\nu_1)+\phi^2]
+\alpha(b+\nu_1)\phi\Sigma^*-\alpha\gamma S_T\phi\Sigma^*,\nonumber\\
&&a_2=\beta\gamma (\phi+2b+2\nu_1)+\alpha\gamma\phi\Sigma^*,\nonumber\\
&&a_3=\beta\gamma^2.\nonumber
\end{eqnarray}
The sequence of signs of the coefficients $a_i$ is either $(-,-,+,+)$ or
$(-,+,+,+)$. There is precisely one change of sign and thus by Descartes' 
rule of signs the polynomial has precisely one positive root. Once a value
of $S$ is given the values of $C_1$ and $C_2$ at the steady state can be 
determined successively. Following the arguments in the case $N=1$ we see 
that $S<S_T$, $C_1+C_2<\Sigma^*$ and that the steady state is unique.

In the case $N=3$ the system is
\begin{eqnarray}
&&\dot S=\alpha C_1(S_T-S)-\beta S,\label{N31}\\
&&\dot C_1=\phi\Sigma^*-(2\phi+b+\gamma S+\nu_1)C_1
+(-\phi+b+\gamma S)C_2-\phi C_3,
\label{N32}\\
&&\dot C_2=\phi C_1-(\phi+b+\gamma S+\nu_1)C_2+(b+\gamma S)C_3,\label{N33}\\
&&\dot C_3=\phi C_2-(b+\gamma S+\nu_1)C_3.\label{N34}
\end{eqnarray}
A calculation for $N=3$ analogous to those already done gives a quartic 
polynomial. Its coefficients are
\begin{eqnarray}
&&a_0=-[(b+\nu_1)^2+\phi\nu_1]\alpha\phi\Sigma^* S_T,\nonumber\\
&&a_1=\beta\gamma\{(\phi+b+\nu_1)[(b(b+\nu_1)+\nu_1(\phi+b+\nu_1)]
+\nu_1(\phi+b+\nu_1)\nonumber\\
&&+\phi^2(b+\nu_1)+\phi^3\}
+[(b+\nu_1)^2+\nu_1\phi]\alpha\phi\Sigma^*
-2(b+\nu_1)\alpha\gamma\phi\Sigma^* S_T,\nonumber\\
&&a_2=\beta\gamma\{b(b+\nu_1)+\nu_1(\phi+b+\nu_1)
+2(\phi+b+\nu_1)(b+\nu_1)+\phi\nu_1+\phi^2\}\nonumber\\
&&+2(b+\nu_1)\alpha\gamma\phi\Sigma^*
-\gamma^2\alpha\phi\Sigma^* S_T,\nonumber\\
&&a_3=\beta\{2\gamma (b+\nu_1)+\gamma^2(\phi+b+\nu_1)\}
+\gamma^2\alpha\phi\Sigma^*,\nonumber\\
&&a_4=\beta\gamma^3.\nonumber
\end{eqnarray}
The coefficient $a_0$ is negative while $a_3$ and $a_4$ are positive. Unless
$a_1>0$ and $a_2<0$ Descartes' rule of signs implies that the polynomial
only has one positive root. Otherwise the rule implies that it has one or
three positive roots (counting multiplicity) but does not decide between 
these two cases.

It will now be shown that in the case $N=3$ there are values of the 
coefficients for which the polynomial $p(S)$ has three positive roots. To do 
this we vary the coefficients $S_T$ and $\nu_1$ in the system 
(\ref{N31})-(\ref{N34}) and keep all others fixed. Note that these coefficients
come from the parameters in the agonist-only case of 
(\ref{phen1})-(\ref{phen4}). To obtain the desired variation of the coefficients
we fix all parameters in (\ref{phen1})-(\ref{phen4}) except $S_T$, $\nu_1$ and 
$\kappa$ and vary $\kappa$ in such a way that $\frac{\nu_1}{\kappa}$ does not
change. This ensures that $\Sigma^*$ does not change. In fact we may simplify 
the calculations by setting $b=0$ since if three positive roots can be obtained
in that case the same thing can be obtained for $b$ small and positive by
continuity. Suppose that $S_T$ and $\nu_1$ depend on a parameter $\epsilon$
with both of them being positive for $\epsilon>0$. Suppose in addition that
in the limit $\epsilon\to 0$ we have the asymptotic relations 
$S_T=\bar S_T\epsilon^{-1}+o(\epsilon^{-1})$ and 
$\nu_1=\bar\nu_1\epsilon^4+o(\epsilon^4)$ for constants $\bar S_T$ and
$\bar\nu_1$. Then we obtain asymptotic expansions
$a_4=A_4$, $a_3=A_3+o(1)$, $a_1=A_1+o(1)$ for positive constants $A_4$, 
$A_3$ and $A_1$, $a_0=A_0\epsilon^3+o(\epsilon^3)$ for a constant $A_0<0$ and  
$a_2=A_2\epsilon^{-1}+o(\epsilon^{-1})$ for a constant $A_2<0$. Let 
$q(S)=\epsilon p(S)$. Then $q(1)$ converges to $A_2$ for $\epsilon\to 0$
and is thus negative for $\epsilon$ small enough. The same is true for
$p(1)$. On the other hand 
\begin{equation}
p(\epsilon^2)=A_0\epsilon^3+A_1\epsilon^2+A_2\epsilon^3
+A_3\epsilon^6+A_4\epsilon^8+o(\epsilon^2)=A_1\epsilon^2+o(\epsilon^2).
\end{equation}
Hence for $\epsilon$ sufficiently small $p(\epsilon^2)>0$. Putting these facts
together shows that $p$ has three positive roots when $\epsilon$ is small. For
each of these roots the values of $C_1$, $C_2$ and $C_3$ at the steady state 
can be determined successively. $S<S_T$, $C_1+C_2+C_3<\Sigma^*$ and defining 
$C_0=\Sigma^*-(C_1+C_2+C_3)$ gives a steady state of the original system.

It has already been noted that $p$ cannot have more than three positive roots. 
There are parameter values for which the positive steady
state is unique. To see this it is enough to assume that $S_T$ is small
while keeping the other parameters fixed. Then $a_i>0$ for all $i>0$ and the 
polynomial can have no more that one positive root since its derivative has
no positive root. These results can be summed up as follows:

\noindent
{\bf Theorem 1} The agonist-only case of the system (\ref{phen1})-(\ref{phen7})
has exactly one positive steady state for $N=1$ and $N=2$. In the case $N=3$ 
there are parameters for which it has three positive steady states and it can 
never have more than three. 

\begin{figure}[ht]
\begin{center}
  \mbox{%
    \includegraphics[scale=0.5]{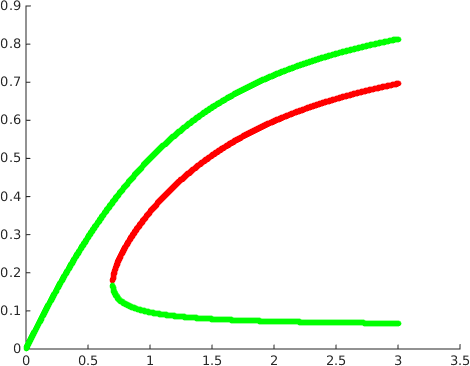}%
\ \ \ 
\includegraphics[scale=0.5]{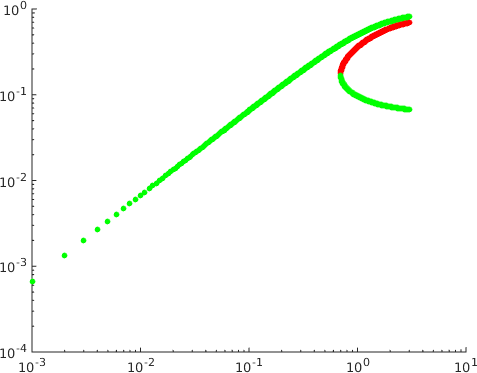}  
  }
\end{center}
\caption{%
Multistability of steady states as a function of $L_1$.  Shown is the coordinate
$C_3$, but other coordinates behave similarly.
Stable branches are shown in green colour and unstable in red.
  Left: linear scale, right: log-log scale.
Parameters are
$\alpha =  1$,
$S_T =  10$,
$\beta =  1$,
$\kappa = 2\times10^{-4}$,
$R =  1$,
$b =   0$, 
$\gamma =  1$,
$\phi = 1$,
$\nu_1 = 1$.
}
\label{fig:multistability}
\end{figure}

A concrete example of parameters for which there are three positive steady 
states is obtained by setting $\alpha$, $\beta$, $\gamma$, $\phi$, $L_1$ and 
$R$ equal to one and choosing $S_T=10$, $\kappa=2\times 10^{-4}$, 
$\nu_1=10^{-4}$. A computer calculation shows that the coordinates 
$(S^*,C_0^*,C_1^*,C_2^*,C_3^*)$ of the steady states are approximately 
\begin{eqnarray}
&&(1.1769,0.1570,0.1334,0.1133,0.0963),\\
&&(0.0005,0.0001,0.0001,0.0003,0.4996),\\ 
&&(0.2860,0.0085,0.0294,0.1028,0.3593). 
\end{eqnarray}
It shows in addition that while the first and second of these steady states
are asymptotically stable the third is a saddle with a one-dimensional 
unstable manifold. A plot of the steady states as a function of the parameter 
$L_1$, see Figure~\ref{fig:multistability}, suggests that there is a fold bifurcation.
For higher values of $N$ it is possible to derive a polynomial equation of 
degree $N+1$ for $S$. There is no obvious reason why this polynomial should
not have an arbitrarily large number of positive roots for $N$ arbitrarily
large. A simple upper bound is that the polynomial can have no more than 
$N$ positive roots for $N$ odd and no more than $N+1$ for $N$ even.

In general it is difficult to obtain information about the stability of the 
steady states by analytic methods. In the case $N=1$ the vector field 
defining the dynamical system has negative divergence and so by Dulac's 
criterion und Poincar\'e-Bendixson theory all solutions converge to the 
steady state as $t\to\infty$. The system can exhibit damped 
oscillations as will now be shown. To do this we choose 
parameters so that 
\begin{equation}\label{balance}
\alpha C_1+\beta=\phi+b+\gamma S+\nu_1.
\end{equation}
For fixed values of the quantities $R$ and $S_T$ the quantities $C_1$ and $S$
are bounded uniformly in the quantities appearing in (\ref{balance}). Thus 
if we make $\alpha$ and $\beta$ small while fixing the other parameters we can 
arrange that the left hand side is smaller than the right hand side. If 
starting from there we make $\beta$ large while fixing the other parameters
we can arrange that the left hand side of (\ref{balance}) is 
greater than the right hand side. It follows that parameter values exist for 
which (\ref{balance}) holds. The reason why this is interesting is that the 
discriminant of the characteristic equation of the linearization is the 
sum of a term which vanishes when (\ref{balance}) holds and the expression
$-4\alpha\gamma (S_T-S)C_1$. Thus when (\ref{balance}) holds the linearization
has eigenvalues with negative real part and non-zero imaginary part and there
are damped oscillations.

An interesting limiting case of the agonist-only system is obtained by assuming
that $\alpha=0$ and $S=0$. We refer to this as the kinetic proofreading 
system since it is closely related to McKeithan's kinetic proofreading model
\cite{mckeithan95}. In fact McKeithan only considered the case $b=0$ but 
this makes no essential difference for the analysis which follows. It was
observed by Sontag \cite{sontag01} that the deficiency zero theorem of chemical
reaction network theory can be applied to McKeithan's system to conclude that
there is a unique steady state in each stoichiometric compatibility
class and that this solution is asymptotically stable in its class. Strictly 
speaking chemical reaction network theory is applied to the extended system 
which includes free receptors and free ligand as variables. To show that the 
steady state is globally asymptotically stable it suffices to show that 
there are no $\omega$-limit points on the boundary. That this is the case can 
be proved just as we did for the full system above. The steady state is 
hyperbolic as follows from Appendix C of \cite{feinberg95a}.

Consider now the full agonist-only system. Setting $\alpha=0$ gives a system 
where the kinetic proofreading system is coupled to a system describing the 
decay of $S$. The steady state of the kinetic proofreading system gives rise to 
a steady state of the agonist-only system with $\alpha=0$ which is on 
the boundary of the biologically feasible region and is a hyperbolic sink. 
Denote its coordinates by $(0,C_j^*)$. For $\alpha$ small and positive there 
exists a hyperbolic sink which is a small perturbation of that for $\alpha=0$. 
It must be in the biologically feasible region since $C_1>0$ there and 
equation (\ref{phen1}) would imply that $\dot S>0$ there if $S$ were negative.  
Thus for sufficiently small values of $\alpha$ there exists a 
positive steady state which is a hyperbolic sink 
$(S^*(\alpha),C_j^*(\alpha))$ close to $(0,C_j^*)$. There exists a 
positive number $r$ such that for $\alpha$ sufficiently small, say 
$\alpha\le\alpha_0$, $(S^*(\alpha),C_j^*(\alpha))$ is the only $\omega$-limit 
point of any solution in the open ball of radius $r$ about that steady 
state. 

Let $h(C_j)$ be the Lyapunov function in the proof of the deficiency 
zero theorem. It is known from general arguments that $\dot h\le 0$ with 
equality only for $C_j=C_j^*$. It follows that on the complement of the ball 
of radius $r$ about the steady state the function $\dot h$ has a
strictly negative maximum. We can consider the behaviour of the function $h$ 
for solutions of the system for positive $\alpha$. For
small $\alpha$ it is still a Lyapunov function on the complement of a small
ball about the steady state while there are no $\omega$-limit points 
except the steady state itself within that ball. Hence for $\alpha$ 
sufficiently small a solution can have no $\omega$-limit points other than 
the steady state. It follows that for $\alpha$ small the steady state 
is globally asymptotically stable. Of course this means that the 
limiting system obtained from the agonist-only system by passing to a solution
through an $\omega$-limit point also has a unique steady state which is 
globally asymptotically stable for $\alpha$ sufficiently small. A similar
argument applies in the case of the full system (\ref{phen1})-(\ref{phen7})
since in that case the system obtained by setting $\alpha$ and $S$ to zero 
is just the product of two copies of the corresponding system in the 
agonist-only case.

\section{The response function}\label{response}

This section is concerned with the agonist-only system. From a biological 
point of view the essential input parameters to the system are the ligand
concentration $L_1$ and the binding time of the ligand to the receptor, 
which in the model corresponds to $\nu_1^{-1}$. The latter is a measure of
the signal strength. The essential output is the value of $C_N$ which is 
a measure of the activation of the T cell. Given values of $L_1$, $\nu_1$ 
and the other parameters we can consider the value of $C_N$ in a steady 
state. In fact it is more convenient to use the quantities $\log C_N$ and
$\log L_1$. This leads to a response function $\log C^*_N=F(\log L_1,\nu_1)$. 
If there is more than one steady state for a given choice of the 
parameters this has to be thought of as a multi-valued function. It might 
naively be thought that $F$ should be an increasing function of $L_1$ and a 
decreasing function of $\nu_1$: more antigen leads to more activation of the T 
cell and a longer binding time leads to more activation. This turns out not to 
be the case and the function $F$ is not a monotone function of its arguments. 
This was observed in the case of the dependence on $L_1$ in the simulations of
\cite{francois13}. It is possible to understand intuitively how this 
situation can arise. An increase in the stimulation of the T cell leads to 
activation of SHP-1 and that in turn has a negative effect on the activation 
of the T cell. Many of the calculations in this section are guided by those in 
\cite{francois13}. 

The behaviour of the response function will be estimated in various parameter
ranges. In order to do this it is useful to parametrize the solutions in a 
certain manner which will now be described. In the case of a steady state the 
equation (\ref{phen3}) is a linear difference equation for the $C_j$ with 
constant coefficients. This suggests looking for power-law solutions, an
idea which motivates the following result.

\noindent
{\bf Lemma 2} Steady state solutions of equations (\ref{phen2})-(\ref{phen4})
in the agonist-only case can be parametrized in the form 
\begin{equation}\label{powerlaw}
C_j=a_+r_+^j+a_-r_-^j
\end{equation}
where the coefficients 
$r_{\pm}$ and $a_{\pm}$ are positive and depend on $S$. The quantities $r_+$ and
$r_-$ are given by 
\begin{equation}
r_{\pm}=\frac{\phi+b+\gamma S+\nu_1\pm\sqrt{(\phi+b+\gamma S+\nu_1)^2
-4\phi(b+\gamma S)}}{2(b+\gamma S)}\label{rpm}
\end{equation}
and satisfy $r_-<1<r_+$.

\noindent
{\bf Proof} Note first that the quantities $r_{\pm}$ in (\ref{rpm}) are
the roots of the characteristic equation 
\begin{equation}\label{char}
\phi+(b+\gamma S)r^2-(\phi+b+\gamma S+\nu_1)r=0
\end{equation}
associated to the difference equation already mentioned and it is obvious
that they are positive. The fact that they satisfy the characteristic equation
is equivalent to the condition that the $C_j$ defined by (\ref{powerlaw}) 
satisfy the steady state form of equation (\ref{phen3}). That $r_-<1<r_+$ can 
be seen by noting that the function on the left hand side of (\ref{char}) is 
negative at $r=1$. The condition that the quantities $C_j$ satisfy the equations
(\ref{phen2})-(\ref{phen4}) with $\dot C_j=0$ is equivalent to the conditions
that they satisfy (\ref{powerlaw}) with $r_{\pm}$ as in (\ref{rpm}) and certain 
coefficients $a_-$ and $a_+$ together with the equations obtained by 
substituting (\ref{powerlaw}) into the equations $\dot C_0=0$ and 
$\dot C_N=0$. The explicit form of these last equations is
\begin{eqnarray}
&&[(b+\gamma S)r_--(\phi+\nu_1)]a_-+[(b+\gamma S)r_+-(\phi+\nu_1)]a_+
=-\nu_1\sum_{j=0}^NC_j\label{c0eq}\\
&&r_-^{N-1}[\phi-(b+\gamma S+\nu_1)r_-]a_-+r_+^{N-1}[\phi-(b+\gamma S+\nu_1)
r_+]a_+=0.\label{c1eq}
\end{eqnarray}
It follows from the discussion in Section \ref{steady} that $\sum_{j=0}^NC_j$, 
which was denoted there by $\Sigma_1^*$, is uniquely determined for fixed values
of the parameters in $(\ref{phen2})-(\ref{phen4})$ and fixed $S$. Thus for 
fixed values of these parameters and $S$ all quantities in (\ref{c0eq}) and 
(\ref{c1eq}) except $a_-$ and $a_+$ are known. It will now be shown that these 
equations have a unique solution for $a_-$ and $a_+$. Note that
\begin{equation}\label{product}
[\phi-(b+\gamma S+\nu_1)r_-][\phi-(b+\gamma S+\nu_1)r_+]=
-\frac{\phi^2\nu_1}{b+\gamma S},
\end{equation}
as can most easily be seen by multiplying out the left hand side of this
equation and substituting for $r_+r_-$ and $r_++r_-$, which are the sum and 
product of the roots of the characteristic equation (\ref{char}). Thus equation 
(\ref{c1eq}) gives a positive expression for $a_+/a_-$. Note also that 
(\ref{product}) implies that the factors in the product on the left hand side
of that equation have opposite signs. Since $r_-<r_+$ the first factor is 
positive and the second negative. Substituting the expression for $a_+/a_-$ 
into (\ref{c0eq}) gives an equation of the form 
\begin{equation}\label{ABeq}
a_-[A-B(r_-/r_+)^{N-1}]=-\nu_1\Sigma_1^*[\phi-(b+\gamma S+\nu_1)r_+]
\end{equation}
whose right hand side is positive. Here
\begin{eqnarray}
&&A=[(b+\gamma S)r_--(\phi+\nu_1)][\phi-(b+\gamma S+\nu_1)r_+]\\
&&B=[(b+\gamma S)r_+-(\phi+\nu_1)][\phi-(b+\gamma S+\nu_1)r_-]
\end{eqnarray}
It follows from the fact that the first factor on the left hand side of 
(\ref{product}) is positive that the first factor in the expression for $A$ is
negative and hence that $A$ itself is positive. In addition, a straightforward
computation shows that $A>B$. If $B$ were 
not positive then the quantity in square brackets on the left hand side of 
(\ref{ABeq}) would be positive. If $B$ is positive then the fact that 
$r_-<r_+$ implies that the quantity in square brackets is again positive. 
Hence in any case (\ref{ABeq}) can be solved to give a unique positive value 
of $a_-$. Then $a_+$ can be determined in such a way that (\ref{c0eq}) and 
(\ref{c1eq}) hold. This completes the proof of Lemma 2.

Lemma 2 shows that for fixed parameters in (\ref{phen2})-(\ref{phen4}) and a 
fixed value of $S$ the steady state values of all the $C_j$ are determined but 
this does not yet give expressions for the $C_j$ which can be directly applied 
to study the properties of the response function. For the purposes of what
follows it is convenient to rewrite (\ref{total1}) in the form
\begin{equation}\label{ctotaleq}
\kappa (L_1-\sum_{j=0}^NC_j)(R-\sum_{j=0}^N  C_j)-\nu_1\sum_{j=0}^N C_j=0.
\end{equation}
The equation for $S$ can be solved to give the relation 
$S=S_T\frac{C_1}{C_1+C_*}$ with $C_*=\frac{\beta}{\alpha}$. Summing the
expression for $C_j$ given in Lemma 2 over $j$ gives
\begin{equation}
\sum_{j=0}^N C_j=a_+\frac{r_+^{N+1}-1}{r_+-1}+a_-\frac{r_-^{N+1}-1}{r_--1}.
\end{equation}
The following equation relating $a_-$ and $a_+$ is equation (21) of 
\cite{francois13}. 
\begin{equation}\label{aplusminus}
a_+=-a_-\left(\frac{r_-}{r_+}\right)^{N+1}\frac{r_+-1}{r_--1}.
\end{equation}
Combining the last two equations gives
\begin{equation}\label{sumcj}
\sum_{j=0}^NC_j=\frac{a_-}{1-r_-}\left[1-\left(\frac{r_-}{r_+}\right)^{N+1}
\right].
\end{equation}

Having completed the necessary preliminaries we now proceed to study the 
qualitative behaviour of the reponse function in different regimes. When $L_1$ 
is small it is to be expected that the concentration of the 
phosphatase is small and that the response function resembles that of the 
kinetic proofreading model. It will now be shown that when $L_1$ is small
the leading term in the function $F$ depends linearly on $\log L_1$ with slope 
one and the additive constant in this linear function will be determined. The 
equation (\ref{ctotaleq}) can be written in the form
\begin{equation}
\sum_{j=0}^NC_j=\frac{\kappa RL_1}{\kappa R+\nu_1}\left[1+
\frac{L_1}{R}\left((\sum_{j=0}^NC_j/L_1)^2
-(\sum_{j=0}^N  C_j/L_1)\right)\right].
\end{equation}
Note that $\sum_{j=0}^N  C_j\le L_1$ so that this equation implies that
\begin{equation}\label{sumleading}
\sum_{j=0}^NC_j=\frac{\kappa RL_1}{\kappa R+\nu_1}(1+qL_1/R).
\end{equation}
where $-\frac14<q<0$. Using (\ref{aplusminus}) it is possible to write down an 
explicit expression for $C_N$, namely
\begin{equation}
C_N=\frac{a_-r_-^N(r_+-r_-)}{r_+(1-r_-)}.
\end{equation}
It follows from (\ref{sumcj}) that
\begin{equation}
C_N=r_-^N\frac{1-\frac{r_-}{r_+}}{1-(\frac{r_-}{r_+})^{N+1}}\sum_{j=0}^N C_j.
\end{equation}
Combining these equations gives
\begin{equation}
C_N=\left\{
r_-^N\frac{1-\frac{r_-}{r_+}}{1-(\frac{r_-}{r_+})^{N+1}}
\frac{\kappa R}{\kappa R+\nu_1}\right\}L_1(1+qL_1/R).
\end{equation}
The function of $r_-$ and $r_+$ in this equation defines a function of $S$.
This function of $S$ tends to a positive limiting value as $S\to 0$. Now
$C_1\le\sum_{j=0}^N C_i=O(L_1)$ and $S=O(C_1)$. Hence for $R$ fixed we can 
replace the function of $r_+$ and $r_-$ in the above expression by its 
limiting value for $S\to 0$. If the resulting relation is plotted 
logarithmically it gives a straight line of slope one as the leading order 
approximation in the limit $\log L_1\to -\infty$.

Next we look at an intermediate regime where the amount of activated SHP-1 
is well away from both zero and $S_T$. As a first step, we obtain an estimate 
for $r_-$ which is sharper than that in Lemma 2. To do this we compute the 
left hand side of the characteristic equation (\ref{char}) for 
$r=\frac{\phi}{\phi+\nu_1}$. The result is 
$-\frac{\phi\nu_1(b+\gamma S)}{(\phi+\nu_1)^2}<0$. It follows that 
$r_-<\frac{\phi}{\phi+\nu_1}$. Hence $1-r_->\frac{\nu_1}{\phi+\nu_1}$. 
Substituting this into (\ref{sumcj}) gives 
$a_->\frac{\nu_1}{\phi+\nu_1}\left(\sum_{j=0}^NC_j\right)$. Note that
$\frac{S}{S_T}\ge\min\left\{\frac{C_1}{2C_*},\frac12\right\}$. Hence a positive
lower bound for $C_1$ implies a positive lower bound for $\frac{S}{S_T}$.

Next we will derive a lower bound for $\gamma S$ in the case that $S_T$ is 
large. This will be proved by contradiction. Suppose that $\gamma S\le\rho$ 
for some $\rho>0$. Then it follows from the characteristic equation that
$r_-\ge\frac{\phi}{\phi+\rho+\nu_1}$. Using this in the equation for $C_1$ gives
$C_1\ge\frac{\phi\nu_1}{(\phi+\nu_1)(\phi+\rho+\nu_1)}
\left(\sum_{j=0}^NC_j\right)$. 
It follows that
\begin{equation}\label{smin}
S\ge S_T\min\left\{\frac{\phi\nu_1}{2C_*(\phi+\nu_1)(\phi+\rho+\nu_1)}
\left(\sum_{j=0}^NC_j\right),\frac12\right\}
\end{equation}
It is then clear that for a given value of $\rho$ and fixed values of the 
parameters other than $S_T$ this leads to a contradiction if $S_T$ is 
sufficiently large. In other words, given any $\rho>0$ there is a lower bound 
for $S_T$ which implies that $\gamma S\ge\rho$. It is convenient to make the 
restrictions that $\kappa R\ge 1$ and $L_1/R\le 1$ since then it is
possible to replace $\sum_{j=0}^NC_j$ in (\ref{smin}) by $\frac{3L_1}{4(1+\nu_1)}$
by using (\ref{sumleading}).

From (\ref{rpm}) it can be concluded that
\begin{eqnarray}
&&r_-=\frac{\phi}{b+\gamma S}(1+O(\eta)),\\
&&r_+=1+O(\eta).
\end{eqnarray}
where $\eta=\frac{\phi+\nu_1}{b+\gamma S}$.
This gives approximate expressions for the roots of the characteristic equation
if $\frac{\phi+\nu_1}{b+\gamma S}$ is small. As a consequence of these 
equations
\begin{equation}
\frac{r_-}{r_+}=\frac{\phi}{b+\gamma S}(1+O(\eta)).
\end{equation}

Taking the expression for $C_1$ supplied by Lemma 2 and using 
(\ref{aplusminus}), (\ref{sumcj}) and (\ref{sumleading}) gives
\begin{equation}
C_1=r_-\frac{\kappa RL_1}{\kappa R+\nu_1}(1+O(\eta)).
\end{equation}
This implies that $C_1=O(\eta)$ and the
expression relating $S$ and $C_1$ then shows that 
$\frac{S}{S_T}=O(\eta)$.
In fact 
\begin{equation}
C_1=\frac{C_*S}{S_T}(1+O(\eta))
\end{equation}

These relations indicate that in leading order $r_-$ is proportional to $S$.
However it is also the case that
\begin{equation}
r_-=\frac{1}{S}\frac{\phi}{\gamma}\frac{1}{1+b/(\gamma S)}(1+O(\eta))
\end{equation}
which indicates that in leading order $r_-$ is proportional to $S^{-1}$.
Hence
\begin{equation}
r_-=\frac{C_*(\kappa R+\nu_1)}{\kappa RL_1S_T}S(1+O(\eta))
\end{equation}
and 
\begin{equation}
r_-=\frac{1}{S}\frac{\phi}{\gamma}(1+O(\eta')).
\end{equation}
where $\eta'=\max\{\eta,b/(\gamma S)\}$.
Combining these two relations gives
\begin{equation}
S=\sqrt{\frac{\phi\kappa RS_TL_1}{C_*\gamma (\kappa R+\nu_1)}}(1+O(\eta')).
\end{equation}
Substituting this back into the equation for $r_-$ gives
\begin{equation}
r_-=\sqrt{\frac{\phi C_*(\kappa R+\nu_1)}
{\gamma S_TL_1\kappa R}}(1+O(\eta')).
\end{equation}
This means that
\begin{eqnarray}
&&C_N=(\sum_{j=0}^N C_j)r_-^N(1+O(\eta''))\nonumber\\
&&=\left(\frac{\kappa R+\nu_1}{\kappa RL_1}\right)^{N/2-1}
\left(\frac{\phi C_*}{\gamma S_T}\right)^{N/2}(1+O(\eta''))\nonumber\\
&&=\left(\frac{\phi \beta}{\alpha\gamma S_T}\right)^{N/2}
\left(\frac{\kappa R+\nu_1}{\kappa R}\right)^{N/2-1}(L_1)^{1-N/2}(1+O(\eta''))
\end{eqnarray}
where $\eta''=\max\{\eta',L_1/R\}$. Choosing $L_1$ small enough makes 
$L_1/R$ small. With $L_1$ fixed, making $S_T$ large enough makes $\eta$ small. 
Thus $\eta''$ can be made as small as desired by choosing
$L_1$ sufficiently small and $S_T$ sufficiently large.

\noindent
{\bf Theorem 2} Consider the response function $\log C_N=F(\log L_1,\nu_1)$ 
for the steady states of the system (\ref{phen1})-(\ref{phen4}) with $L_2=0$
and $D_j=0$. Choose fixed values for all parameters in the system except 
$L_1$ and $S_T$. Suppose that $\kappa R\ge 1$. Let $\epsilon>0$. Then there 
exists a constant $\delta$ with $0<\delta\le R$ such 
that the following holds. If $0<L_0<\delta$ there exists $\mu>0$ such that if 
$S_T\ge\mu$ the inequality 
\begin{equation}
\left|\left(\frac{\phi \beta}{\alpha\gamma S_T}\right)^{-N/2}
\left(\frac{\kappa R+\nu_1}{\kappa R}\right)^{1-N/2}(L_1)^{N/2-1}
F(\log L_1,\nu_1)-1\right|<\epsilon 
\end{equation}
holds on the interval $[\log L_0,\delta]$.

\noindent
{\bf Proof} To obtain the conclusion of the theorem it suffices to show 
that under the given assumptions $\eta''$ can be made as small as desired. 
That this is possible follows from the discussion above.

Note that this theorem implies, in particular, that for $N>2$ and suitable 
values of $L_1$ and $S_T$ there exists a range of $L_1$ in which the 
response function is decreasing. The theorem also implies that in this 
regime the reponse function can be an increasing function of $\nu_1$. This 
effect was not captured by the calculations of \cite{francois13} since there
$\frac{\nu_1}{\kappa R}$ was assumed to be so small as to be negligible.

Finally we examine the regime where $L_1/R$ is small but the phosphatase is 
close to being completely activated. This means that $S/S_T$ is
close to one. This holds provided $C_1$ is sufficiently large compared to 
$C_*$. It remains to check that such a regime actually occurs for some 
values of the parameters. It is possible to make $\sum_{j=0}^NC_j$ large 
while keeping $L_1/R$ constant. This can be done by making $R$ large.
This makes $a_-$ large without making $r_-$ small. Hence it makes $C_1$ large
and hence $S$ close to $S_T$. In this regime the function of $r_+$ and $r_-$
occurring in the expression for $C_N$ can be replaced by its limit for 
$S\to S_T$ and we again get a region where the slope of the graph of 
$\log C_N$ as a function $\log L_1$ is one but the line has been shifted
compared to that obtained for $L_1/R$ small.

\begin{figure}[ht]
\begin{center}
\includegraphics[scale=0.5]{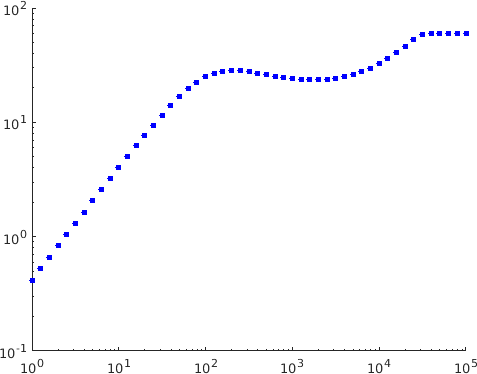}
\end{center}
\caption{%
Log-log plot showing linearity of $\log C_3$ as a function of $\log L_1$ for
small $L_1$, followed by decreasing, increasing, and saturation regimes.
Parameters are
$\alpha =  1$,
$S_T =  6\times10^5$,
$\beta =  5\times10^2$,
$\kappa = 10^{-4}$,
$R =  3\times10^4$,
$b = 4\times10^{-2}$,
$\gamma =  1.2\times10^{-6}$,
$\phi = 9\times10^{-2}$,
$\nu_1 = 10^{-2}$.
}
\label{fig:regimes}
\end{figure}

In \cite{francois13} these types of behaviour were exhibited numerically in
the case $N=5$ with biologically reasonable choices of the parameters. We
found that changing these parameters a little allows similar observations to
be made in the case $N=3$.  In the plot shown in Figure~\ref{fig:regimes}
the three regimes can be seen 
together with a fourth regime where $L_1/R$ is no longer small. It is clear 
that a regime of this type must exist since the response function is globally 
bounded.

\begin{figure}[ht]
\begin{center}
  \mbox{%
    \includegraphics[scale=0.5]{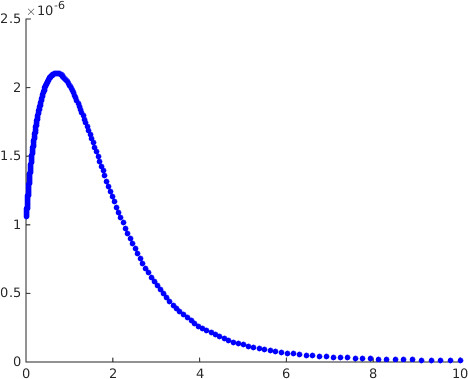}%
\ \ \ 
\includegraphics[scale=0.5]{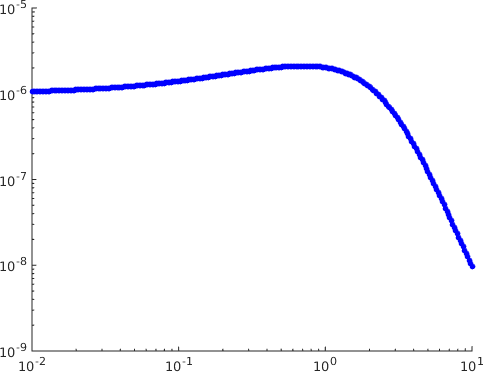}
}
\end{center}
\caption{%
  $C_3$ as a function of $\nu_1$ in model with $N=3$, showing non-monotonic
  behaviour for some values of parameters.
  Left: linear scale, right: log-log scale.
Parameters are
$\alpha =  10^{-1}$, 
$S_T =  10^7$, 
$\beta =  10$,
$\kappa = 10^{-6}$, 
$R =  10^5$, 
$b =   10^{-2}$, 
$\gamma =  10^{-4}$, 
$\phi = 10^{-2}$, 
$L_1 = 10^3$. 
}
\label{fig:c3nu1}
\end{figure}

We now turn to the dependence of the response function on $\nu_1$.
It has been suggested in \cite{lever14} that the kinetic proofreading model 
with negative feedback as studied here is not able to explain the presence of 
an optimal dissociation time, a biological effect confirmed by the 
experimental work of \cite{lever16}. The plots of the response as a function 
of the dissociation time in that type of model in \cite{lever14} show that it 
is increasing. Having an optimal dissociation time would require that there be 
a region where this function is decreasing. The response function being 
increasing as a function of the dissociation time corresponds to its being 
decreasing a function of $\nu_1$. Here we have given an analytical proof in 
Theorem 2 that there exist parameters for which the response is an increasing 
function of $\nu_1$, in contrast to the plots in \cite{lever14}. Since the 
theorem is of limited help in finding explicit parameters for which 
this happens we also did a numerical search and identified parameters
of this type. The results are displayed in Figure~\ref{fig:c3nu1}, where it is seen that 
$F$ has a maximum as a function of $\nu_1$ for fixed $L_1$, which 
corresponds to an optimal dissociation time.
The conclusion of both the 
analytical and the numerical work is as follows. The claim that the kinetic 
proofreading model with feedback can only produce a response which is a 
decreasing function of the parameter $\nu_1$ is dependent on the parameter 
values chosen to do the simulations and not a general property of the model.

\section{Including the antagonist}\label{antagonist}

When the antagonist is included the output variable expressing the degree of
activation of the T cell is $C_N+D_N$. Now asymptotic expressions for this
quantity will be derived. It has already been shown that for a steady state of 
the system (\ref{phen1})-(\ref{phen7}) the quantities $\sum_{j=0}^NC_j$ and 
$\sum_{j=0}^ND_j$ can be expressed in terms of the parameters. The equation for 
$S$ can be solved to give the relation $S=S_T\frac{C_1+D_1}{C_1+D_1+C_*}$. $C_j$ 
solves the same difference equation as in the agonist-only case and $D_j$ 
solves the difference equation obtained from that one by replacing $\nu_1$ by 
$\nu_2$. The quantities $r_-$, $r_+$, $a_-$ and $a_+$ differ in the two cases. We
can nevertheless proceed as in the former case to see that the solutions
for $C_j$ and $D_j$ allow parametrizations in terms of these quantities as
before. Note that using the equations (\ref{total1}) and (\ref{total2}) 
it is possible to eliminate the $D_j$ from the equation for $C_0$ and the 
$C_j$ from the equation for $D_0$. Thus we have coupled equations for the 
$C_j$ and $D_j$ which can be analysed just as in the agonist-only case to
express $C_1$ and $D_1$ as functions of $S$ and the parameters. We can also
write $C_N$ and $D_N$ as functions of $\Sigma_1$ and $\Sigma_2$ respectively.
Proceeding as in the agonist-only case we get an expression for $C_N+D_N$ in 
the kinetic proofreading regime. The multiple of $L_1$ obtained there as 
leading term is replaced by a linear combination of $L_1$ and $L_2$.

Next the intermediate regime will be considered. For this it is necessary to
define a new parameter $\eta=\max\{\frac{\phi+\nu_1}{b+\gamma S}\}$. There are 
asymptotic expressions for $r_-$ and $r_+$ where the leading terms are just as 
in the agonist-only case. In particular they are the same for $C_j$ and $D_j$.
Two asymptotic expressions for the quantity $C_1+D_1$ can be obtained.
\begin{eqnarray}
&&C_1+D_1=\frac{C_*S}{S_T}(1+O(\eta)),\\
&&=r_-\left(\frac{\kappa RL_1}{\kappa R+\nu_1}
+\frac{\kappa RL_2}{\kappa R+\nu_2}\right)(1+O(\eta)).
\end{eqnarray}
This gives an expression for $r_-$ in terms of $S$. As in the agonist-only
case this gives an expression for $r_-$ where the dependence on $S$ has been
eliminated in leading order.
\begin{equation}
r_-=\sqrt{\frac{\phi C_*}{\gamma S_T}
\left(\frac{\kappa RL_1}{\kappa R+\nu_1}
+\frac{\kappa RL_2}{\kappa R+\nu_2}\right)^{-1}}(1+O(\eta’))
\end{equation}
where $\eta'$ is defined in terms of $\eta$ as in the agonist-only case.
Following the steps used in the agonist-only case leads to an expression
for $C_N+D_N$ which is the same as that previously obtained for $C_N$ except
that the expression $\frac{\kappa RL_1}{\kappa R+\nu_1}$ is replaced by
$\frac{\kappa RL_1}{\kappa R+\nu_1}+\frac{\kappa RL_2}{\kappa R+\nu_2}$.
This leads in the end to an asymptotic expression for $C_N+D_N$ under a 
suitable assumption $L_1$ and $L_2$. The assumption made in the 
agonist-only case can naturally be written as an assumption on 
$\frac{\kappa RL_1}{\kappa R+\nu_1}$ and in the present case it is replaced
by an assumption on 
$\frac{\kappa RL_1}{\kappa R+\nu_1}+\frac{\kappa RL_2}{\kappa R+\nu_2}$.
This implies that under certain circumstances $C_N+D_N$ increases when 
$L_2$ increases and $L_1$ is held fixed. An increase in the amount of 
self-antigen can lead to a decrease in the response to a foreign antigen.

\section{Conclusions and outlook}

In this paper some properties of the solutions of the model of \cite{francois13}
for T cell activation were proved. A new discovery was that already in the 
case of three phosphorylation sites ($N=3$) there can exist more than one 
positive steady state for given values of the parameters. Another new 
observation is that damped oscillations can occur. It was also proved 
that, as suggested by the calculations in \cite{francois13}, the output 
variable $C_N$ (concentration of the maximally phosphorylated receptor) is
a decreasing function of the concentration $L_1$ of antigen in some 
parts of parameter space. In an analogous way it was proved that under
some circumstances the activation in response to an agonist can be decreased
by increasing the concentration of the antagonist. It was proved that it can
also happen that $C_N$ is an increasing function of the dissociation
constant $\nu_1$. This abstract result was given a concrete illustration
by a plot showing that $C_N$ can have a local maximum as a function of $\nu_1$. 

The stability of the steady 
states was only determined analytically in the very special cases $N=1$ and 
$\alpha$ close to zero. For $N=3$ numerical calculations showed the 
occurrence of two stable steady states for certain values of the parameters.
It was proved that damped oscillations occur but can there
also be sustained oscillations (periodic solutions)? It is thus clear
that there remain several aspects of the dynamics of this system which would 
profit from further investigations, analytical and numerical.

In immunology it is important to describe diverse situations including the 
course of different types of infectious disease, the development of autoimmune
diseases and the destruction of tumour cells by the immune system. It would be 
unreasonable to expect that a simple mechanism could be the key to describing 
all these situations. One strategy to try to obtain more understanding is to
choose one mechanism and to investigate which types of situations it suffices 
to describe. This may be done by combining mathematical models with experimental
data. What are the restrictions under which the type of model studied in this
paper might be appropriate? The first assumption is that in the situation to
be explained the distinction between self and non-self takes place within an
individual T cell. In other words it is assumed that it is not necessary to 
consider the population dynamics of the T cells involved or even the interaction
of their population with that of other types of immune cells such as 
regulatory T cells or dendritic cells. A quite different type of mathematical
model, where population effects are considered, can be found in 
\cite{sontag17}. 
In that case, in contrast to the lifetime dogma, the response depends on the 
rate of change of the antigen concentration. The second assumption which is 
important for the models studied here is that the distinction between self and 
non-self takes place on a sufficiently short time scale, say three minutes. On 
longer time scales there may be essential effects related to the spatial 
distribution of molecules on the T cell surface (formation of the 
immunological synapse) so that a description by means of ordinary differential 
equations may be insufficient. It may also happen that some T cell receptors 
become 
inactive on a longer time scale (limiting signalling model, cf. \cite{lever16}).

In this paper we have concentrated on studying the mathematical properties
of a particular model for the biological phenomenon of T cell activation
with arbitrary values of the parameters. A complementary question is to
what extent known experimental data on the parameters may further 
constrain the dynamics in this model. In addition it is important to know
whether this model is consistent with all biological data and how it 
compares to other possible models for the same biological system. For a 
discussion of this we refer to \cite{lever14}, \cite{francois15} and
\cite{lever16}. It was indicated in \cite{lever16} that the situation
where $C_N$ is a decreasing function of $\nu_1$ cannot be reproduced using
the model of \cite{francois13}. Our results indicate that a failure of the
model to reproduce this effect must depend not only on the model itself
but on the choice of parameters used for simulations. At the same time it 
may be that this effect only occurs in experiments where the measurements 
are done on long time scales (many hours) and not on the time scale of 
the initial activation (a few minutes) for which the models of 
\cite{altanbonnet05} and \cite{francois13} were primarily intended.
We plan to investigate these questions further elsewhere.

\end{document}